\documentclass[12pt]{article}

\usepackage{amssymb,amsmath}
\usepackage{latexsym}

\usepackage[T2A]{fontenc} 
\usepackage[cp1251]{inputenc} 
\newtheorem{theorem}{Theorem}
\newtheorem{corollary}{Corollary}
\newtheorem{lemma}{Lemma}
\newtheorem{proposition}{Proposition}

\begin{document}

\begin{center}
{\bf Tensor fields defined by
Lax representations}
\end{center}

\begin{center}
Alexander Vladimirovich Balandin  \\
Department of Algebra, Geometry and
Discrete Mathematics \\
N.I.Lobatchevsky Nizhny Novgorod State University\\
23 Gagarin ave., 603950 Nizhny Novgorod, Russia    \\
e-mail: balandin@mm.unn.ru
\end{center}

\begin{abstract}

In this paper, some properties of tensor fields constructed by
the Lax representation of chiral-type systems are investigated.

\end{abstract}



\section{Introduction}

Chiral-type systems (see, for example, \cite{DMM}) are the systems
of partial differential equations of the form
\begin{equation} \label{EQ1}
\Delta^\alpha \equiv U^{\alpha}_{xy} +
G^{\alpha}_{\beta\gamma}U^{\beta}_xU^{\gamma}_y +Q^{\alpha} = 0.
\end{equation}
Here and further, the Greek indices 
range from 1 to $n$
and the subscripts denote partial derivatives with respect to the
independent variables $x$ and $y$. The coefficients
$G^{\alpha}_{\beta\gamma},Q^{\alpha}$ are assumed to be smooth
functions of the variables  $U^1, U^2, ..., U^n$. The summation rule
over the repeated indices is also assumed.

Further on, the covariant derivatives w.r.t. the connection
defined by the coefficients $G^{\alpha}_{\beta\gamma}$
are denoted by $\nabla_\delta.$

The Euler-Lagrange equations of the form (\ref{EQ1}) are called
a nonlinear generalized sigma model.

Following \cite{Olver}, recall that the characteristic of the  conservation law $L=(L_1,L_2)$
of the system  (\ref{EQ1}) is
a set of functions $R = \{R_\alpha\}$ such that
$$Div\; L= D_xL_1+D_yL_2=  R_\alpha \Delta^\alpha ,$$
where $\Delta^\alpha $ denotes the l.h.s. of Eq. (\ref{EQ1}).

Characteristics of the  conservation laws are also referred to as cosymmetries.

We understand integrable systems as the systems admitting  a Lax representation.

In the sequel, we  assume  that the system (\ref{EQ1}) admits  
the matrix $\mathfrak{g}$-valued
Lax representation of the form:
\begin{equation}\label{EQ2}
D_y\widetilde{A}-D_x\widetilde{B}+[\widetilde{A},\widetilde{B}]=S_\alpha \Delta^\alpha,
\end{equation}
where

\begin{equation}\label{EQ3}
\widetilde{A} =A_\alpha U^\alpha_x +\lambda M,\;\;\; \widetilde{B}=B_\alpha U^\alpha_y +\frac{1}{\lambda}N,
\end{equation}
\begin{equation}\label{EQa3}
S_\alpha =A_\alpha -B_\alpha,
\end{equation}
 $ A,B,M,N$ are smooth functions of the variables $U^1,U^2,...,U^n$, taking values in
a matrix Lie algebra $\mathfrak{g}.$

It is easy to see, collecting the terms by $ \lambda, \frac{1}{\lambda},$
that $M,N, A_\alpha,B_\alpha$ satisfy the following conditions

\begin{equation}\label{EQ5}
M_{,\alpha}=[ B_\alpha,M],
\end{equation}
\begin{equation}\label{EQ6}
 N_{,\alpha}=[ A_\alpha,N].
\end{equation}

\begin{equation}\label{EQ61}
Q^\alpha S_\alpha=\frac{1}{2}[M,N],
\end{equation}

\begin{equation}\label{EQ1s}
A_{\alpha,\beta}-B_{\beta,\alpha}+[A_\alpha,B_\beta]-S_\gamma G^\gamma_{\alpha\beta}=0.
\end{equation}

Here and further, comma denotes the partial derivatives,
that is, $P_{,\alpha}=\frac{\partial P}{\partial U^\alpha}.$

The set of such functions $S_\alpha$ is referred to as  a characteristic element of
the Lax representation \cite{Mar}. In the sequel, it is assumed that $S_1,S_2,...,S_n $ are linear independent.

\textbf{Remark 1} Note that the functions $M,N,$ and the characteristic element $S_\alpha$ of  the Lax representation (\ref{EQ2},\ref{EQ3},\ref{EQa3})  are transformed under a gauge transformation by the formulas

$$M\rightarrow T^{-1}MT,\; N\rightarrow T^{-1}NT,\;S_\alpha\rightarrow T^{-1}S_\alpha T.$$

Thus, the functions of the form
$f(S_{\alpha_1},S_{\alpha_2},...S_{\alpha_p},\underbrace{M,M,...,M}_r,\underbrace{N,N,...,N}_s) $
are well defined tensor fields for an arbitrary ad-invariant symmetric $(p+r+s)$-form on $\mathfrak{g}.$
It turns out that these tensor fields carry important
information about the system under consideration.

It was mentioned in \cite{Bal2} that for an arbitrary ad-invariant
symmetric $(k+p)$-form on $\mathfrak{g},$ tensor fields of the form
$$F_{\alpha_1\alpha_2...\alpha_k}=f(S_{\alpha_1},S_{\alpha_2},...S_{\alpha_k},\underbrace{M,M,...,M}_p),\;
(k> 0,\;p\ge 0)$$
are Killing fields, that is
$$\nabla_{(\beta }F_{\alpha_1\alpha_2...\alpha_k)}=0.\; 
$$
Here the proof of it is given (Proposition 1).

The important case $k=1$ of such tensors was considered in \cite{Bal1}. It turns out that
the sets $F_{\alpha}$ 
defined by the expressions
\begin{equation}\label{EQ4}
 F_{\alpha}=f(S_\alpha, M,M,...M)
\end{equation}
are  characteristics of the zero order of  conservation laws (cosymmetries of the zero order) of the system (\ref{EQ1}).

In this paper, in a similar way, we investigate
the meaning  of the tensors
$$K_{\alpha_1\alpha_2...\alpha_k}=f ([S_{\alpha_1},M],[S_{\alpha_2},M],...,[S_{\alpha_k},M],M,...,M).
$$
It is shown that covariant derivatives of
these tensors vanish (Theorem 1). According to well known result,  these tensors for $k=2$ define conservation
laws $K_{\alpha\beta}U^\alpha_x U^\beta_xdx
$ for the 
system
\begin{equation} \label{REQ1}
\Delta^\alpha \equiv U^{\alpha}_{xy} +
G^{\alpha}_{\beta\gamma}U^{\beta}_xU^{\gamma}_y  = 0,
\end{equation}
that is, the sets $K_{\alpha\beta}U^\alpha_x 
$ form the
cosymmetries of the first order for system \eqref{REQ1}.

Note that the sets $K_{\alpha\beta}U^\alpha_x 
$ do not form the
cosymmetries  for the general case of system \eqref{EQ1} with
non vanishing $Q^\alpha$
(Example 1).

However, it is valid for  3-dimensional Lie algebras.
This result was
announced in \cite{Bal1}. Here the full proof of it is given (Theorem 2).

\textbf{Remark 2} It is obvious that the 
results obtained from
the above mentioned ones are valid also if we change  $x\leftrightarrow y,\;M\leftrightarrow N,\;
G^{\alpha}_{\beta\gamma}\leftrightarrow G^{\alpha}_{\beta\gamma}-2G^{\alpha}_{[\beta\gamma]}.$

\medskip
\section{Main theorems }

\begin{proposition}
Let the system \eqref{EQ1} admit
the matrix $\mathfrak{g}$-valued Lax representation of the form (\ref{EQ3}) and $f$ be a
symmetric ad-invariant $p$-form $f$ on Lie algebra $\mathfrak{g}$. 
Then the  tensor field
\begin{equation}
\label{kf1}
F_{\alpha_1\alpha_2...\alpha_k}=f(S_{\alpha_1},S_{\alpha_2},...S_{\alpha_k},\underbrace{M,M,...,M}_p),\;
(k> 0,\;p\ge 0)
\end{equation}
satisfy the condition
\begin{equation}
\label{kf2}\nabla_{(\beta }F_{\alpha_1\alpha_2...\alpha_k)}=0,\; 
\end{equation}
that is, $F_{\alpha_1\alpha_2...\alpha_k}$ is a Killing field. 

\end{proposition}

\textbf{Proof}.

Rewrite Eq.\eqref{EQ1s} in the form
\begin{equation}
\label{t11}
\nabla_\beta S_\alpha =S_{\alpha,\beta}-S_\gamma G^\gamma_{\alpha\beta}=[B_\beta,S_\alpha]+D_{\alpha\beta},
\end{equation}
where
\begin{equation}\label{d1}
D_{\alpha\beta}=[B_\beta,B_\alpha]+2B_{[\beta,\alpha]}.
\end{equation}
Then, one can obtain

$$\nabla_{\beta }F_{\alpha_1\alpha_2...\alpha_k}=
f(\nabla_{\beta }S_{\alpha_1},S_{\alpha_2},...S_{\alpha_k},\underbrace{M,M,...,M}_p)
$$
$$+
f(S_{\alpha_1},\nabla_{\beta }S_{\alpha_2},...S_{\alpha_k},\underbrace{M,M,...,M}_p)
$$
$$+...+f(S_{\alpha_1},S_{\alpha_2},...,\nabla_{\beta }S_{\alpha_k},\underbrace{M,M,...,M}_p)+
pf(S_{\alpha_1},S_{\alpha_2},...S_{\alpha_k},\underbrace{M_{,\beta},M,...,M}_p).
$$
Using Ad-invariancy of form $f$ and Eq.\eqref{EQ5}, Eq.\eqref{t11}, one can see that
\begin{equation}
\label{t12}
\nabla_{\beta }F_{\alpha_1\alpha_2...\alpha_k} =f(D_{\alpha_1\beta},S_{\alpha_2},...S_{\alpha_k},\underbrace{M,M,...,M}_p)
\end{equation}
$$+
f(S_{\alpha_1},D_{\alpha_2\beta},...S_{\alpha_k},\underbrace{M,M,...,M}_p)
$$
$$f(S_{\alpha_1},S_{\alpha_2},...D_{\alpha_k \beta},\underbrace{M,M,...,M}_p).
$$
Symmetrizing Eq.\eqref{t12} and taking into account that $D_{(\alpha\beta)}=0,$ we finish the proof.
\bigskip

\bigskip
\bigskip

Denote the $\alpha-$th Euler operator  by
$$E_\alpha= \sum_{J}(-D)_J(\frac{\partial}{\partial U^\alpha_J}),$$
where the sum extending over all multi-indices $J=(j_1,j_2).$

To proceed we need  the following lemma.

\begin{lemma}
Let $R$  be a function of the form:
\begin{equation}\label{1EQ7}
R=
K_{\alpha \beta}U^{\alpha}_{xy}U^{\beta}_{x} +
L_{\alpha\beta\gamma}U^{\alpha}_xU^{\beta}_xU^{\gamma}_y
+W_{\alpha}U^{\alpha}_x,
\end{equation}
where $K_{\alpha \beta},L_{\alpha\beta\gamma},
W_{\alpha}$ are functions of $U^1,U^2,...,U^n.$

Then,
the equations
$$
E_{\mu}(R)=0$$
 are equivalent to the following conditions:
\begin{equation}\label{2EQ7}
K_{[\alpha \mu]}=0,
\end{equation}
\begin{equation}\label{3EQ7}
K_{\mu\alpha ,\beta} -2L_{(\mu\alpha)\beta} =0,
\end{equation}
\begin{equation}\label{4EQ7}
W_{[\alpha,\mu]}=0.
\end{equation}
Here and further, the comma denotes the partial derivatives,
that is, $P_{,\alpha}=\frac{\partial P}{\partial U^\alpha}.$

\end{lemma}

\textbf{Proof}.

The proof is obtained by direct computation. Collecting the terms by $U^\alpha_{xxy},U^\alpha_{xy}U^\beta_y,
U^\alpha_{x}U^\beta_xU^\gamma_y,U^\alpha_x$ in the expression $E_\mu(R)$ and
taking into account the condition $E_{\mu}(R)=0 ,$ we
obtain Eq.(\ref{2EQ7}),(\ref{3EQ7}), the equation $K_{\alpha\beta,[\mu,\gamma]}=0$ (which is an identity),
and  Eq.(\ref{4EQ7}).

\textbf{Remark 3} Assuming that
\begin{equation}\label{L21}
L_{\alpha \beta\gamma}=K_{\alpha \mu}G^\mu_{\beta\gamma},\; W_\alpha=K_{\alpha \beta}Q^\beta.
\end{equation}
Then one can easily verify that Eq. \eqref{4EQ7} can be rewritten in the
form
\begin{equation}\label{7EQ7}
 (Q^\alpha K_{\alpha \beta})_{,\gamma}-(Q^\alpha K_{\alpha \gamma})_{,\beta}=0,
\end{equation}
and  Eq. \eqref{2EQ7}, \eqref{3EQ7} are equivalent to the condition
\begin{equation}\label{61EQ7}
\nabla_\gamma K_{\alpha \beta}=0.
\end{equation}

\medskip

\medskip

\begin{theorem}

Let the system \eqref{EQ1} admit
the matrix $\mathfrak{g}$-valued Lax representation of the form (\ref{EQ3}) and $f$ be a
symmetric ad-invariant $p$-form $f$ on Lie algebra $\mathfrak{g}$. 

Then the tensor field
\begin{equation}
\label{r}
K_{\alpha_1\alpha_2...\alpha_k}=f ([S_{\alpha_1},M],[S_{\alpha_2},M],...,[S_{\alpha_k},M],\underbrace{M,...,M}_{p-k})
\end{equation}

satisfies the condition 

\begin{equation}
\label{b} \nabla_\gamma K_{\alpha_1\alpha_2...\alpha_k}=0.
\end{equation}

\end{theorem}

We will give the proof of the theorem in Appendix.

\medskip
\medskip

\begin{corollary}

Let the system \eqref{REQ1} 
admit
the matrix $\mathfrak{g}$-valued Lax representation of the form (\ref{EQ3}). 
Then for every  symmetric ad-invariant $p$-form $f$ on Lie algebra $\mathfrak{g}$ 
the sets

$$
Y_\alpha=f([S_\alpha,M], [S_\beta,M],\underbrace{M,M,...,M}_{p-2})U^\beta_x $$
form cosymmetriies of the first order of the system under consideration.

\end{corollary}

The proof follows from the well known result that $K_{\alpha\beta}U^\alpha_x U^\beta_xdx$ is an integral
of system \eqref{REQ1} iff $\nabla_\gamma K_{\alpha \beta}=0. $ 
\medskip
\medskip

\begin{corollary}
Let $n$-component  system \eqref{REQ1} 
admit
$\mathfrak{g}$-valued Lax representation of the form (\ref{EQ3}), where $\mathfrak{g}$ is compact algebra Lie
of rank $l$ and $l> n$. Assume that
$M
$ is a regular
element of Lie algebra $\mathfrak{g}$ in a point $P_0.$

Then the  covariant constant tensor field $K_{\alpha\beta}$ defined by
\begin{equation}\label{k33}
K_{\alpha\beta}=f([S_\alpha,M], [S_\beta,M]),
\end{equation}
where $f$ is a Killing form on $\mathfrak{g},$
does not vanish at point $P_0.$

\end{corollary}

\textbf{Proof}. One can obtain the proof taking into account the linear independence
of $S_\alpha,$ the compactness of $\mathfrak{g},$ and reasons of the dimensionality.

\medskip
\medskip

\medskip

\textbf{Remark 4}. It is well known that
the Euler-Lagrange system  for Lagrangian
$$L= g_{\alpha\beta}(U^\gamma)U^\alpha_xU^\beta_y+Q(U^\gamma), $$
where $ g_{[\alpha\beta]}=0, det||g_{\alpha\beta} || \ne0,$ admits the
integrals $g_{\alpha\beta}U^\alpha_xU^\beta_xdx,\; g_{\alpha\beta}U^\alpha_yU^\beta_ydy. $
Thus, if tensor field $f([S_\alpha,M], [S_\beta,M],\underbrace{M,M,...,M}_{p-2}) $ is up to a constant proportional
to
the metric $ g_{\alpha\beta}$ determined by Lagrangian $L,$ then this tensor field
forms a cosymmetry. Note that no examples of such systems admitting the Lax representation with dim $\mathfrak{g}>3$
are known to the author.

\textbf{Remark 5}.

Corollary 1 could not be
generalized to the case 
 of the chiral-type systems \eqref{EQ1} with non vanishing $Q^\alpha$.
This illustrates the following example.

\medskip
\medskip

\medskip

\textbf{Example 1}. Consider the 3-component variational system

$$\Delta^1=U^1_{xy} -aU^2e^{2U^1}+bU^3e^{-2U^1}=0,$$
$$\Delta^2=
U^2_{xy} -b\psi^{-1}e^{-2U^1} -\psi U^3U^2_xU^2_y=0 ,
$$
$$\Delta^3=U^3_{xy} -a\psi^{-1}e^{2U^1} -\psi U^2U^3_xU^3_y=0,
 $$
where $\psi = (U^2U^3+c)^{-1}$ and $a,b,c$ are  arbitrary constants, and $a^2+b^2\ne 0.$

This system is the Euler system for Lagrangian
\begin{equation}
\label{L1}
L= 2U^1_xU^1_y+ \psi(U^2_xU^3_y+U^3_xU^2_y) +2aU^2e^{2U^1}+2bU^3e^{-2U^1}
\end{equation}
and 
admits the Lax representation which takes values in $sl(3),$
where \cite{DMM}:

$$
\widetilde{A}=\left (
\begin{array}{ccc}
-\frac{U^2_xU^3}{3(U^2U^3+c)}&-bc^{-1}U^3e^{-2U^1}&0\\
\lambda aU^2e^{2U^1}&-\frac{U^2_xU^3}{3(U^2U^3+c)}&\lambda ae^{2U^1}\\
0&bc^{-1}e^{-2U^1}(U^2U^3+c)&\frac{2U^2_xU^3}{3(U^2U^3+c)}
\end{array}
\right ),
$$
$$
\widetilde{B}=\left (
\begin{array}{ccc}
-\frac{1}{3}(2U^1_y+\frac{U^2U^3_y}{U^2U^3+c})& \lambda^{-1}&-\frac{U^3_y}{U^2U^3+c}\\
-c&\frac{1}{3}(4U^1_y-\frac{U^2U^3_y}{U^2U^3+c})&0\\
-U_y^2&0&-\frac{1}{3}(2U^1_y-\frac{2U^2U^3_y}{U^2U^3+c})
\end{array}
\right ).
$$

This Lax representation is not of the form (\ref{EQ3}).
In order to construct tensor $K_{\alpha\beta}$ using Eq.\eqref{k33},
we change the Lax representation to the form:
$$
\widetilde{A}=\left (
\begin{array}{ccc}
-\frac{U^2_xU^3}{3(U^2U^3+c)}&-bc^{-1}U^3e^{-2U^1}\lambda &0\\
\lambda aU^2e^{2U^1}&-\frac{U^2_xU^3}{3(U^2U^3+c)}&\lambda ae^{2U^1}\\
0&bc^{-1}e^{-2U^1}(U^2U^3+c)\lambda&\frac{2U^2_xU^3}{3(U^2U^3+c)}
\end{array}
\right ),
$$
$$
\widetilde{B}=\left (
\begin{array}{ccc}
-\frac{1}{3}(2U^1_y+\frac{U^2U^3_y}{U^2U^3+c})& \lambda^{-1}&-\frac{U^3_y}{U^2U^3+c}\\
-c\lambda^{-1}&\frac{1}{3}(4U^1_y-\frac{U^2U^3_y}{U^2U^3+c})&0\\
-U_y^2&0&-\frac{1}{3}(2U^1_y-\frac{2U^2U^3_y}{U^2U^3+c})
\end{array}
\right ).
$$

It means that this representation  admits two spectral parameters.

Choosing $f(x,y)$ as Killing form $tr(xy),$ one can find by direct calculation

\begin{equation}
\label{k20}
K =-ab\left (
\begin{array}{ccc}
8&0&0\\
0&0&\frac{1}{U^2U^3+c}\\
0&\frac{1}{U^2U^3+c}& 0
\end{array}
\right ).
\end{equation}

Analogously one can obtain
by direct calculations:
$$\widetilde{K}= 
f([S_\alpha,N], [S_\beta,N])=
c\left (
\begin{array}{ccc}
4&0&0\\
0&0&\frac{1}{U^2U^3+c}\\
0&\frac{1}{U^2U^3+c}&0
\end{array}
\right )$$

Consider the case $a=b=0.$ Then, one can easily see that $\widetilde{K}_{\alpha\beta}U^\alpha_y
$ is
the characteristic
of the conservation law $c[4(U^1_y)^2+ \frac{U^2_yU^3_y}{U^2U^3+c}]dy$, in accordance with corollary 1.

Note that in general case $a\ne0$ or $b\ne0$ matrices $K,\widetilde{K}$ are not proportional up to a constant to
the metric defined by Lagrangian \eqref{L1}, and one can verify that the sets $\widetilde{K}_{\alpha\beta}U^\beta_x,\;K_{\alpha\beta}U^\beta_x$ do not form
a cosymmetry of the system under consideration.

It interesting to note that the linear combination
$cK +3ab\widetilde{K} $ is proportional to the metric and define a cosymmetry.

\bigskip

\medskip
\medskip

\medskip

\begin{theorem} Let the system \eqref{EQ1} admit the Lax representation
of the form (\ref{EQ3}) valued in a 3-dimensional Lie algebra.
Then
 for every  symmetric ad-invariant $p$-form $f$ on Lie algebra $\mathfrak{g}$ the set
\begin{equation}\label{84EQ7}
Y_\alpha=f([S_\alpha,M], [S_\beta,M],\underbrace{M,M,...,M}_{p-2})U^\beta_x,\;
\end{equation}
forms cosymmetry of the first order of the system under consideration (\ref{7EQ7}).

\end{theorem}

\textbf{Proof}. The proof for the two cases of $so(3)$ and $sl(2)$ is the same due
to the well known isomorphism over $\mathbb{C}$ of these Lie algebras. The only statement we need to prove is the condition (\ref{7EQ7}).
Denote $Z_\beta =Q^\alpha K_{\alpha \beta}.$ Then, using Eq. (\ref{EQ61}), we have the following equalities
$ Z_\beta =f([Q^\alpha S_\alpha,M], [S_\beta,M],\underbrace{M,M,...,M}_{p-2})\\
= \frac{1}{2}f([[M,N],M], [S_\beta,M],\underbrace{M,M,...,M}_{p-2})\\ = -\frac{1}{2}f(M(M,N), [S_\beta,M],\underbrace{M,M,...,M}_{p-2})+\frac{1}{2}f(N(M,M), [S_\beta,M],\underbrace{M,M,...,M}_{p-2}),$
where the round brackets denote the scalar product w.r.t. the Killing metric on $\mathfrak{g}.$
Now, one can see that the first summand vanish due to Ad-invariance of the form $f.$
Thus, we obtain
\begin{equation}\label{86EQ7}
Z_\beta = \frac{1}{2}(M,M)f(N, [S_\beta,M],\underbrace{M,M,...,M}_{p-2})\\ =
\end{equation}
$$-\frac{1}{2}(M,M)f([N,M], S_\beta,\underbrace{M,M,...,M}_{p-2}). $$
It turns out that the last equation can be integrated.
At first, note that from condition (\ref{EQ5}) it follows that $(M,M)=const.$
Denote by $H$ the function
$H=f(N,\underbrace{M,M,...,M}_{p-1}).$ Now, one can verify, using Eq.(\ref{EQ5}) and (\ref{EQ6}), the following identity:
$H_{,\alpha}=\frac{\partial H}{\partial U^\alpha}=pf([N,M],S_\alpha,\underbrace{M,M,...,M}_{p-2}). $
Then, we arrive at $ Z_\beta =-\frac{1}{2p}(M,M)H_{,\beta}$ and the proof is finished.
\medskip

\textbf{Example 3}.
Pohlmeier-Lund-Regge system \cite{LR},

 $$
\Delta^1=  U^1_{xy}+\frac{1}{\sin U^2}(U^1_xU^2_y+U^1_yU^2_x)=0,
 $$
  $$
\Delta^2= U^2_{xy}-\frac{\sin U^2}{(1+\cos U^2)^2}U^1_xU^1_y-p \sin U^2=0,
 $$
where $p$ is an arbitrary constant.
It will be convenient to write the Lax representation of PLR system in  form (\ref{EQ2}),(\ref{EQ3}),
where
$$
\widetilde{A}=\left (
\begin{array}{ccc}
0&\lambda p-\frac{\cos U^2 U^1_x }{2\cos^2 \frac{U^2}{2}}&-tg \frac{U^2}{2}U^1_x\\
-(p\lambda-\frac{\cos U^2 U^1_x }{2\cos^2 \frac{U^2}{2}})&0&U^2_x \\
tg \frac{U^2}{2}U^1_x&-U^2_x&0
\end{array}
\right ),
$$
$$
\widetilde{B}=\left (
\begin{array}{ccc}
0&-(\frac{\cos U^2}{\lambda}+\frac{U^1_y}{2\cos^2 \frac{U^2}{2}})&-\frac{\sin U^2}{\lambda}\\
\frac{\cos U^2}{\lambda}+\frac{U^1_y}{2\cos^2 \frac{U^2}{2}}&0&0\\
-\frac{\sin U^2}{\lambda}&0&0
\end{array}
\right ).
$$
Choose the following basis $B$ of the Lie algebra $\mathfrak{so(3)}$
$$
\overrightarrow{e}_1 =\left (
\begin{array}{ccc}
0&0&0\\
0&0&1\\
0&-1&0
\end{array}
\right ),\overrightarrow{e}_2=\left (
\begin{array}{ccc}
0&0&-1\\
0&0&0\\
1&0&0
\end{array}
\right ),\overrightarrow{e}_3=\left (
\begin{array}{ccc}
0&1&0\\
-1&0&0\\
0&0&0
\end{array}
\right );
$$
then  we have

\begin{eqnarray*}
D_y\widetilde{A}-D_x\widetilde{B}+[\widetilde{A},\widetilde{B}]=S_\alpha \Delta^\alpha=
  \left (
\begin{array}{ccc}
0&tg^2\frac{U^2}{2}&-tg\frac{U^2}{2}\\
-tg^2\frac{U^2}{2}&0&0\\
tg\frac{U^2}{2}&0&0
\end{array}
\right )\Delta^1
\\+\left (
\begin{array}{ccc}
0&0&0\\
0&0&1\\
0&-1&0
\end{array}
\right )\Delta^2.
\end{eqnarray*}
Thus, w. r. t.
the basis $B$,
$$
S_1=(0,tg\frac{U^2}{2},tg^2\frac{U^2}{2}),\;S_2=(1,0,0),$$
$$M=(0,0,p),\;N=(0, \sin U^2,-\cos U^2). $$

Assume that the 2-form $f$ is the Killing metric of the Lie algebra $\mathfrak{so}(3)$ which, w. r. t.  the  basis
$B$,  is $\delta_{ij}$ up to a constant factor.

Construct a characteristic of the first order, using expressions \eqref{84EQ7}. Assuming
that $f$ is the Killing form, we obtain
$$K=f([S_\alpha,M], [S_\beta,M]) = p^2\left (
\begin{array}{cc}
\tan ^2\frac{U^2}{2}&0\\
0&1\\
\end{array}
\right ),$$
$$\widetilde{K}_{\alpha_\beta}=f([S_\alpha,N], [S_\beta,N]) = \left (
\begin{array}{cc}
\tan ^2\frac{U^2}{2}&0\\
0&1\\
\end{array}
\right ) ,$$
and
$$Y_1=p^2U^1_x\tan ^2\frac{U^2}{2},\; Y_2=p^2U^2_x,\; \widetilde{Y}_1=U^1_y\tan ^2\frac{U^2}{2},\;
 \widetilde{Y}_2=U^2_y. $$
One can verify that $Y$ and $\widetilde{Y}$ are
 the characteristics of the following
conservation laws: $p^2[(U^1_x)^2\tan ^2\frac{U^2}{2}+(U^2_x)^2]dx-2p^2\cos U^2dy $ and
$[(U^1_x)^2\tan ^2\frac{U^2}{2}+(U^2_y)^2]dy-2\cos U^2dx $, respectively.

\textbf{Example 4}.  Consider the 3-component system

$$U^1_{xy} + U^3_xU^1_y ctgU^3 -\frac{1}{\sin U^3}U^3_yU^2_x +
= 0, $$
$$
U^2_{xy} + U^3_yU^2_x ctgU^3 -\frac{1}{\sin U^3}U^3_xU^1_y
= 0,
$$
$$U^3_{xy} + U^1_yU^2_x \sin U^3
- p\sin U^3 
= 0, $$
where $p$ is an arbitrary constant.
This system
admits the Lax representation of the form (\ref{EQ2}),(\ref{EQ3}),
where \cite{Bal3},\cite{Bal1}:

$$
\widetilde{A}=\left (
\begin{array}{ccc}
0&i\lambda M^3&-i\lambda M^2\\
-\lambda M^3&0&i\lambda M^1 \\
i\lambda M^2&-i\lambda M^1&0
\end{array}
\right ),
$$
$$
\widetilde{B}=\left (
\begin{array}{ccc}
0&\frac{i}{\lambda}-(\cos U^3 U^1_y + U^2_y)&-b_{31}\\
-\frac{i}{\lambda}+(\cos U^3 U^1_y + U^2_y)&0&b_{23}\\
b_{31}&-b_{23}&0
\end{array}
\right ),
$$
$$M^1 = p\sin U^3\sin U^2,\;
M^2 = -p\sin U^3 \cos U^2 ,\;
M^3 = p\cos U^3, $$
$$
b_{31}=\sin U^3\cos U^2 U^1_y - \sin U^2U^3_y,\;\;
b_{23}=-\cos U^2 U^3_y - \sin U^2 \sin U^3U^1_y.
$$
Consider the same basis $B$ and the same 2-form f as in example 1. Then, we find
$S_1=(\sin U^2\sin U^3, - \cos U^2\sin U^3, \cos U^3)$,
$S_2=(0,0,1)$, $S_3=(\cos U^2, \sin U^2,0 ).$

Again, using \eqref{84EQ7} and  assuming $f$ to be
the Killing form, one can find
$$ K_{\alpha\beta} =f([S_\alpha,M], [S_\beta,M])= p^2\left(
\begin{array}{ccc}
0&0&0\\
0&\sin^2 U^3&0\\
0&0&1\\
\end{array}\right),
$$
$$ \widetilde{K}_{\alpha\beta} =f([S_\alpha,N], [S_\beta,N])=  -\left(
\begin{array}{ccc}
\sin^2 U^3&0&0\\
0&0&0\\
0&0&1\\
\end{array}\right).
$$

Thus, according to the Theorem 2, the set $ Y_\alpha =p^2(0,\sin^2 U^3 U^2_x, U^3_x)$ is a characteristic
of the conservation law. Indeed, the corresponding  conservation law is
$$\theta= - p^3\cos U^3dy + \frac{p^2}{2}dx[(U^3_x)^2 + (U^2_x \sin U^3)^2].$$

Analogously, one can
see that 
 $\widetilde{Y}_\alpha =-(\sin^2 U^3U^1_y,0 , U^3_y)$ and
corresponding conservation law is
of the form
$$\widetilde{\theta} =  p\cos U^3dx - \frac{dy}{2}[(U^3_y)^2 + (U^1_y \sin U^3)^2].$$

\medskip

\section{Conclusion}

In this paper, the geometric meaning of some tensor fields constructed by
the Lax representation of chiral-type systems is shown.
The formula for cosymmetries of the first order 
for the chiral-type systems admitting
$\mathfrak{g}$-valued Lax representation $(dim\; \mathfrak{g} =3)$ has been proved.

It seems interesting to find similar formulas for cosymmetries of higher orders.

\section{Acknowledgment}

The author is grateful to E.V. Ferapontov, M.I. Kuznetsov, and Y.V.
Tuzov for useful  discussions.
This work was supported by
the project 1410 of Russian Ministry of Education and Science.

\section{Appendix}

Here, we  prove theorem 1.

According to Eq. \eqref{EQ5},\eqref{t11} rewrite
$\nabla_\gamma K_{\alpha_1\alpha_2...\alpha_k}$ by the following way
\begin{equation}
\nonumber
\nabla_\gamma K_{\alpha_1\alpha_2...\alpha_k}=
f([\nabla_\gamma S_{\alpha_1},M],[S_{\alpha_2},M],...,[S_{\alpha_k},M],\underbrace{M,...,M}_{p-k})
\end{equation}
$$+f ([S_{\alpha_1},M_{,\gamma}],[\nabla_\gamma S_{\alpha_2},M],...,[S_{\alpha_k},M],\underbrace{M,...,M}_{p-k})+...
$$
$$+f ([S_{\alpha_1},M],[S_{\alpha_2},M],...,[S_{\alpha_k},M_{,\gamma}],\underbrace{M,...,M}_{p-k}) $$
$$+(p-k)f ([S_{\alpha_1},M],[S_{\alpha_2},M],...,[S_{\alpha_k},M],M_{,\gamma},\underbrace{M,M,...,M}_{p-k-1})$$
$$=\underline{f([[B_\gamma, S_{\alpha_1}],M],[S_{\alpha_2},M],...,[S_{\alpha_k},M],\underbrace{M,...,M}_{p-k})}$$
$$+f([D_{\alpha_1\gamma},M],[S_{\alpha_2},M],...,[S_{\alpha_k},M],\underbrace{M,...,M}_{p-k})
$$
$$+\underline{f([S_{\alpha_1},[B_\gamma, M]],[S_{\alpha_2},M],...,[S_{\alpha_k},M],\underbrace{M,...,M}_{p-k})}
$$
$$+\underline{f([S_{\alpha_1},M],[[B_\gamma,S_{\alpha_2}],M],...,[S_{\alpha_k},M],\underbrace{M,...,M}_{p-k})}$$
$$+ f([S_{\alpha_1},M],[D_{\alpha_2\gamma},M],...,[S_{\alpha_k},M],\underbrace{M,...,M}_{p-k}) $$
$$+ \underline{f([S_{\alpha_1}, M],[S_{\alpha_2},[B_\gamma, M]],...,[S_{\alpha_k},M],\underbrace{M,...,M}_{p-k})}$$
$$+...+ \underline{(p-k)f ([S_{\alpha_1},M],
[S_{\alpha_2},M],...,[S_{\alpha_k},M],[B_{,\gamma},M]\underbrace{M,...,M}_{p-k-1})}.$$

Now, one can see that the underlined terms vanish due to Ad-invariancy of the form $f.$

Next, one can obtain from Eq.\eqref{EQ5} the following identity
\begin{equation}\label{81EQ7}
[2B_{[\alpha,\gamma]},M]=-[[B_\alpha,B_\gamma],M].
\end{equation}
Taking into account Eq.\eqref{d1}, one can find that $[D_{\alpha\beta},M]=0. $
The proof is complete.

\bibliographystyle{amsplain}

\end{document}